\documentclass[aps,prc,showpacs,showkeys,superscriptaddress,footinbib,eqsecnum]{revtex4}
\usepackage{graphicx}
\usepackage{amsmath}
\usepackage{amsfonts}
\usepackage{amssymb}
\usepackage{bm}
\newcommand{\beq}{\begin{equation}}
\newcommand{\eeq}{\end{equation}}
\newcommand{\beqa}{\begin{eqnarray}}
\newcommand{\eeqa}{\end{eqnarray}}
\newcommand{\eq}[1]{Eq.~(\ref{#1})}
\newcommand{\nn}{\nonumber \\ }

\date{\today}
\begin{document}

\title{Peripheral nucleon-nucleon scattering at fifth order of chiral perturbation theory}

\author{D. R. Entem}
\email{entem@usal.es}
\affiliation{Grupo de F\'isica Nuclear, IUFFyM, Universidad de Salamanca, E-37008 Salamanca,
Spain}
\author{N. Kaiser}
\email{nkaiser@ph.tum.de}
\affiliation{Physik Department T39 , Technische Universit\H{a}t M\H{u}nchen, D-85747 
Garching, Germany}
\author{R. Machleidt}
\email{machleid@uidaho.edu}
\affiliation{Department of Physics, University of Idaho, Moscow, Idaho 83844, USA}
\author{Y. Nosyk}
\affiliation{Department of Physics, University of Idaho, Moscow, Idaho 83844, USA}

\begin{abstract}
We present the two- and three-pion exchange contributions to the 
nucleon-nucleon interaction which occur
at next-to-next-to-next-to-next-to-leading order 
(N$^4$LO, fifth order) of chiral effective field theory, and calculate
nucleon-nucleon scattering in peripheral partial waves with $L\geq3$
using low-energy constants that were extracted from $\pi N$ analysis
at fourth order. While the net three-pion exchange contribution is moderate,
the two-pion exchanges turn out to be sizeable and prevailingly repulsive, thus,
compensating the excessive attraction characteristic for NNLO and N$^3$LO.
As a result, the N$^4$LO predictions for the phase shifts of peripheral partial waves are 
in very good agreement with the data (with the only exception of the $^1F_3$ wave).
We also discuss the issue of the order-by-order convergence of the chiral expansion for the
$NN$ interaction. 
\end{abstract}

\maketitle

\section{Introduction}
\label{sec_intro}

During the past three decades, it has been demonstrated that chiral effective field theory 
(chiral EFT) represents a powerful tool to deal with hadronic interactions at low energy in 
a systematic and model-independent way (see Refs.~\cite{ME11,EHM09} for recent reviews). 
The systematics is provided by
a low-energy expansion arranged in terms of powers of the soft scale over the hard scale,
$(Q/\Lambda_\chi)^\nu$,  where $Q$ is generic for an external
momentum (nucleon three-momentum or pion four-momentum) or a pion mass,
and $\Lambda_\chi \approx 1$ GeV the chiral symmetry breaking scale.
The model-independent dynamics is created by pions interacting under the constraint of 
broken chiral symmetry which provides the link to low-energy QCD. 

The early applications of chiral perturbation theory (ChPT) focused on systems like 
$\pi\pi$~\cite{GL84} and $\pi N$~\cite{GSS88}, where the Goldstone-boson character of the 
pion guarantees that a perturbative expansion exists.
But the past 20 years have also seen great progress in applying ChPT to nuclear forces
\cite{ME11,EHM09,Wei90,ORK94,KBW97,KGW98,Kai00a,Kai00b,Kai01,Kai01a,Kai02,EGM98,EM02,EM03,
EGM05}. About a decade ago, the nucleon-nucleon ($NN$) interaction up to fourth order
(next-to-next-to-next-to-leading order, N$^3$LO) was
derived~\cite{KBW97,Kai00a,Kai00b,Kai01a,Kai02,EM02} and quantitative $NN$ potentials
were developed~\cite{EM03,EGM05}.

These  N$^3$LO $NN$ potentials complemented by chiral three-nucleon forces (3NFs) 
 have been applied in
calculations of few-nucleon 
reactions, the structure of light- and medium-mass nuclei, 
and nuclear and neutron matter---with, in general, a good deal of success.
However, some problems continue to exist that seem to defy any solution. 
The most prominent one is the so-called `$A_y$ puzzle' of nucleon-deuteron scattering,
which requires the inclusion of three-nucleon forces (3NFs)~\cite{EMW02}.
While the chiral 3NF at NNLO slightly  improves the predictions for low-energy $N-d$
scattering~\cite{Viv13}, inclusion of the N$^3$LO 3NF deteriorates the 
predictions~\cite{Gol14}. Based upon general arguments, the N$^3$LO 3NF is presumed weak, 
which is why one would not expect the solution of any substantial problems, anyhow.
When working in the framework of an expansion, then, the obvious way to proceed 
is to turn to the next order, which is N$^4$LO (or fifth order). 
Some 3NF topologies at N$^4$LO have already been worked out~\cite{KGE12,KGE13},
and it has been shown that, at this order, all 22 possible isospin-spin-momentum 3NF
structures appear. Moreover, the contributions are moderate to sizeable.
What makes the fifth order even more interesting is the fact that, at this order,
a new set of 3NF contact interactions appears, which has recently been derived
by the Pisa group~\cite{GKV11}. 3NF contact terms are attractive from the point of view 
of the practitioner, because they are typically simple (as compared
to loop contributions) and their coefficients are essentially free. 
Thus, at N$^4$LO, the $A_y$ puzzle may be solved in a trivial way through 
3NF (contact) interactions. Due to the great diversity of structures offered at N$^4$LO,
one can also expect that other persistent nuclear structure problems may finally
find their solution at N$^4$LO.

A principle of all EFTs is that, for reliable predictions, it is necessary that  
all terms included are evaluated at the order at which the calculation is conducted. 
Thus, if nuclear structure problems require for their solution the inclusion of
3NFs at N$^4$LO, then also the two-nucleon force involved in the calculation has 
to be of order N$^4$LO. This is one reason for the investigation of the
$NN$ interaction at N$^4$LO presented in this paper.
Besides this, there are also some more specific issues that motivate a study of this kind.
From calculations of the $NN$ interaction at NNLO~\cite{KBW97} and 
N$^3$LO~\cite{EM02}, it is wellknown that
there is a problem with excessive attraction, particularly, when for the $c_i$
low-energy constants  (LECs) of the dimension-two $\pi N$ Lagrangian the values are 
applied that are obtained from $\pi N$ analysis. It is important to know if this problem 
is finally solved when going beyond N$^3$LO. Last not least, also the convergence of 
the chiral expansion of the $NN$ interaction is of general interest.

This paper is organized as follows:
In Sec.~II, we derive the two- and three-pion exchange contributions at fifth order. 
The predictions
for $NN$ scattering in peripheral partial waves are shown in Sec.~III, and Sec.~IV 
concludes the paper. In the Appendices, we summarize the detailed mathematical expressions 
that define the lower orders of the chiral $NN$ potential. This is necessary, because in this 
study we perform the power counting (of relativistic  $1/M_N$-corrections) differently as 
compared to our earlier work. Since we present also phase shift predictions for the lower 
orders, the unambiguous definition of each order is necessary to avoid confusion.

\section{Pion-exchange contributions to the $NN$ potential}
\label{sec_pions}

The various pion-exchange contributions to the $NN$ potential may be analyzed
according to the number of pions being exchanged between the two
nucleons:
\begin{equation}
V = V_{1\pi} + V_{2\pi} + V_{3\pi} + \ldots \,,
\end{equation}
where the meaning of the subscripts is obvious
and the ellipsis represents $4\pi$ and higher pion exchanges. For each of the above terms, 
we have a low-momentum expansion:
\begin{eqnarray}
V_{1\pi} & = & V_{1\pi}^{(0)} + V_{1\pi}^{(2)} 
+ V_{1\pi}^{(3)} + V_{1\pi}^{(4)} + V_{1\pi}^{(5)} + \ldots 
\label{eq_1pe_orders}
\\
V_{2\pi} & = & V_{2\pi}^{(2)} + V_{2\pi}^{(3)} + V_{2\pi}^{(4)} + V_{2\pi}^{(5)} 
+ \ldots \\
V_{3\pi} & = & V_{3\pi}^{(4)} + V_{3\pi}^{(5)} + \ldots \,,
\end{eqnarray}
where the superscript denotes the order $\nu$ of the expansion, which for an irreducible 
two-nucleon diagram is given by $\nu =  2L + \sum_i (d_i + n_i/2 - 2)$ with $L$ the number 
of loops, $d_i$ is the number of derivatives or pion-mass insertions, and $n_i$ the number 
of nucleon fields (nucleon legs) involved in vertex $i$. The sum runs over all vertices 
contained in the diagram under consideration. 

Order by order, the $NN$ potential builds up as follows:
\beqa
V_{\rm LO} & \equiv & V^{(0)} =
V_{1\pi}^{(0)} 
\label{eq_VLO}
\\
V_{\rm NLO} & \equiv & V^{(2)} = V_{\rm LO} +
V_{1\pi}^{(2)} +
V_{2\pi}^{(2)} 
\label{eq_VNLO}
\\
V_{\rm NNLO} & \equiv & V^{(3)} = V_{\rm NLO} +
V_{1\pi}^{(3)} + 
V_{2\pi}^{(3)} 
\label{eq_VNNLO}
\\
V_{\rm N3LO} & \equiv & V^{(4)} = V_{\rm NNLO} +
V_{1\pi}^{(4)} + 
V_{2\pi}^{(4)} +
V_{3\pi}^{(4)} 
\label{eq_VN3LO}
\\
V_{\rm N4LO} & \equiv & V^{(5)} = V_{\rm N3LO} +
V_{1\pi}^{(5)} + 
V_{2\pi}^{(5)} +
V_{3\pi}^{(5)} 
\label{eq_VN4LO}
\eeqa
where 
LO stands for leading order, NLO for next-to-leading order, etc..

In past work~\cite{ORK94,KBW97,KGW98,Kai00a,Kai00b,Kai01a,Kai02,EGM98,EM02,EM03,EGM05}, 
the $NN$ interaction has been developed up to N$^3$LO.
To make this paper selfcontained and, because we perform the power counting for 
relativistic corrections differently as compared to our previous work, we summarize, order 
by order, the contributions up to N$^3$LO in the Appendices. In this way, all orders, which 
we are talking about in this paper, are unambiguously defined.

The novel feature of this paper are the contributions to the $NN$ potential at N$^4$LO,
which we will present now.

The results will be stated in terms of contributions to the 
momentum-space $NN$ amplitudes in the center-of-mass system (CMS), 
which arise from the following general decomposition:
\begin{eqnarray} 
V({\vec p}~', \vec p) &  = &
 \:\, V_C \:\, + \bm{\tau}_1 \cdot \bm{\tau}_2 \, W_C 
\nonumber \\ & & + 
\left[ \, V_S \:\, + \bm{\tau}_1 \cdot \bm{\tau}_2 \, W_S 
\,\:\, \right] \,
\vec\sigma_1 \cdot \vec \sigma_2
\nonumber \\ &+& 
\left[ \, V_{LS} + \bm{\tau}_1 \cdot \bm{\tau}_2 \, W_{LS}    
\right] \,
\left(-i \vec S \cdot (\vec q \times \vec k) \,\right)
\nonumber \\ &+& 
\left[ \, V_T \:\,     + \bm{\tau}_1 \cdot \bm{\tau}_2 \, W_T 
\,\:\, \right] \,
\vec \sigma_1 \cdot \vec q \,\, \vec \sigma_2 \cdot \vec q  
\nonumber \\ &+& 
\left[ \, V_{\sigma L} + \bm{\tau}_1 \cdot \bm{\tau}_2 \, 
      W_{\sigma L} \, \right] \,
\vec\sigma_1\cdot(\vec q\times \vec k\,) \,\,
\vec \sigma_2 \cdot(\vec q\times \vec k\,)
\, ,
\label{eq_nnamp}
\end{eqnarray}
where ${\vec p}\,'$ and $\vec p$ denote the final and initial nucleon momenta in the CMS, 
respectively. Moreover, $\vec q = {\vec p}\,' - \vec p$ is the momentum transfer, 
$\vec k =({\vec p}\,' + \vec p)/2$ the average momentum, and $\vec S =(\vec\sigma_1+
\vec\sigma_2)/2 $ the total spin, with $\vec \sigma_{1,2}$ and $\bm{\tau}_{1,2}$ the spin 
and isospin operators, of nucleon 1 and 2, respectively.
For on-shell scattering, $V_\alpha$ and $W_\alpha$ ($\alpha=C,S,LS,T,\sigma L$) can be 
expressed as functions of $q= |\vec q\,|$ and $p=|{\vec p}\,'| = |\vec p\,|$, only. Note that 
the one-pion exchange contribution in Eq.~(\ref{eq_1pe_orders}) is of the form 
$W_T^{(1\pi)} = - (g_{\pi N}/2M_N)^2 (m_\pi^2+q^2)^{-1}$ with physical values of the coupling 
constant $g_{\pi N}$ and nucleon and pion masses $M_N$ and $m_\pi$. This expression fixes at 
the same time our sign-convention for $V({\vec p}~', \vec p)$.

We will state two-loop contributions in terms of their spectral functions, from which
the momentum-space amplitudes $V_\alpha(q)$ and $W_\alpha(q)$
are obtained
via the subtracted dispersion integrals:
\begin{eqnarray} 
V_{C,S}(q) &=& 
-{2 q^6 \over \pi} \int_{nm_\pi}^{\tilde{\Lambda}} d\mu \,
{{\rm Im\,}V_{C,S}(i \mu) \over \mu^5 (\mu^2+q^2) }\,, 
\nn
V_T(q) &=& 
{2 q^4 \over \pi} \int_{nm_\pi}^{\tilde{\Lambda}} d\mu \,
{{\rm Im\,}V_T(i \mu) \over \mu^3 (\mu^2+q^2) }\,, 
\label{eq_disp}
\end{eqnarray}
and similarly for $W_{C,S,T}$.
Clearly, $n=2$ for two-pion exchange and $n=3$ for
three-pion exchange.
For $\tilde{\Lambda} \rightarrow \infty$ the above dispersion integrals yield the
results of dimensional regularization, while for finite $\tilde{\Lambda} \geq nm_\pi$
we have what has become known  as spectral-function regularization (SFR) \cite{EGM04}. The 
purpose of the finite scale $\tilde{\Lambda}$ is to constrain the imaginary parts to the  
low-momentum region where chiral effective field theory is applicable.

\subsection{Two-pion exchange contributions at N$^4$LO}
\label{sec_2pi}

\begin{figure}
\scalebox{0.7}{\includegraphics{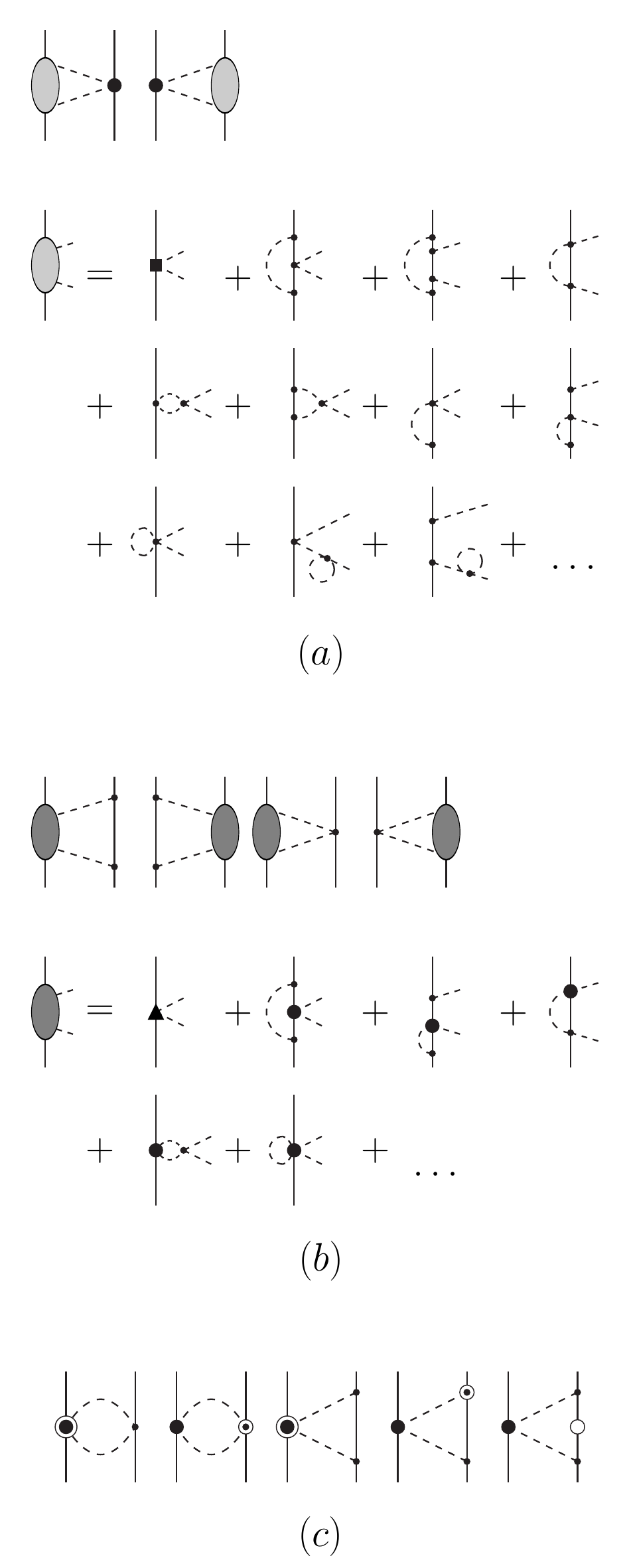}}
\caption{Two-pion-exchange contributions at N$^4$LO.
(a) The leading one-loop $\pi N$ amplitude is folded with the chiral 
$\pi\pi NN$ vertices proportional to $c_i$. 
(b) The
one-loop $\pi N$ amplitude
proportional to $c_i$ 
is folded with the leading order chiral $\pi N$
amplitude.
(c) Relativistic corrections of NNLO diagrams.
Solid lines
represent nucleons and dashed lines pions. 
Small dots, large solid dots, solid squares, and triangles
denote vertices of index $d_i+n_i/2-2 =$ 0, 1, 2, and 3, respectively. 
Open circles  are relativistic $1/M_N$ corrections.
}
\label{fig_dia4}
\end{figure}

The $2\pi$-exchange contributions that occur at N$^4$LO are displayed graphically in 
Fig.~\ref{fig_dia4}. We present now the corresponding analytical expressions separately 
for each class. 

\subsubsection{Spectral functions for class (a)}
\label{sec_2pia}

The N$^4$LO $2\pi$-exchange two-loop contributions of class (a) are shown
in Fig.~\ref{fig_dia4}(a). For this class the spectral functions are obtained by 
integrating the product of the leading one-loop $\pi N$ amplitude and the chiral 
$\pi\pi NN$ vertex proportional to $c_i$ over the Lorentz-invariant $2\pi$-phase space. 
In the $\pi\pi$ center-of-mass frame this integral can be expressed as an angular 
integral $\int_{-1}^1 dx$ \cite{Kai01a}. The results for the non-vanishing spectral 
functions read:
\begin{eqnarray} 
{\rm Im}V_C &\!\!\!\!=\!\!\!\!& - {m_\pi^5\over
(4f_\pi)^6 \pi^2}\Bigg\{
g_A^2 \sqrt{u^2-4}\,\bigg(5-2u^2-{2\over u^2}\bigg)\Big[
24c_1+c_2(u^2-4)+6c_3(u^2-2)\Big]
\ln{u+2 \over u-2}\nonumber \\ && +{8\over
u}\Big[3\big(4c_1+c_3(u^2-2)\big)(4g_A^4u^2
-10g_A^4+1)+c_2 (6g_A^4u^2-10g_A^4-3)\Big] B(u)\nonumber \\ &&
+\sqrt{u^2-4}\,\bigg[3(2-u^2)\big(4c_1+c_3(u^2-2)\big)+c_2(7u^2-6-u^4)+{4g_A^2\over u}
(2u^2-1)\nonumber \\ && \times \Big[4(6c_1-c_2-3c_3)+(c_2+6c_3)u^2\Big]+
4g_A^4
\bigg({32\over u+2}(2c_1+c_3)+{64\over 3u}(6c_1+c_2-3c_3)\nonumber \\ &&
+14c_3-5c_2-92c_1+{8u\over 3}(18c_3-5c_2)+{u^2\over 6}(36c_1+13c_2-156c_3)
+{u^4\over 6}(2c_2+9c_3)\bigg)\bigg] \Bigg\}\,, 
\\
{\rm Im}W_S &=& \mu^2\, {\rm Im}W_T =  {c_4\, g_A^2  
m_\pi^5 \over
(4f_\pi)^6\pi^2}\bigg\{8g_A^2u(5-u^2)B(u)+{1\over 3}(u^2-4)^{5/2}
\ln{u+2 \over u-2} \nonumber
\\ && +{u\over 3} \sqrt{u^2-4}\,\Big[g_A^2(30u-u^3-64)
-4u^2+16\Big]\bigg\}\,,
\end{eqnarray}
with the dimensionless variable $u = \mu/m_\pi>2$ and the logarithmic function
\begin{equation}
B(u) = \ln{u+\sqrt{u^2-4}\over 2}\,.
\end{equation}

\subsubsection{Spectral functions for class (b)}
\label{sec_2pib}
The N$^4$LO $2\pi$-exchange two-loop contributions of class (b) are displayed
in Fig.~\ref{fig_dia4}(b).
For this class, the product of the one-loop $\pi N$ amplitude
proportional to $c_i$ (see Ref.~\cite{KGE12} for details)
and the leading order chiral $\pi N$ amplitude is integrated over the $2\pi$-phase space.
We obtain:
\begin{eqnarray}
{\rm Im}V_S &=& \mu^2\, {\rm Im}V_T = {g_A^4m_\pi^5(c_3-c_4)u\over
(4f_\pi)^6\pi^2}\Big\{\sqrt{u^2-4}\,(u^3-30u+64)+24(u^2-5)B(u) \Big\}\,,
\\
{\rm Im}W_S &=& \mu^2\, {\rm Im}W_T ={g_A^2m_\pi^5\over(4f_\pi)^6 \pi^2}(4-u^2)\Bigg\{{c_4\over
3}\bigg[\sqrt{u^2-4}
\,(2u^2-8)B(u) \nonumber \\ && + 4u(2+9g_A^2)-{5u^3\over 3}\bigg]
+2\bar e_{17} (8\pi f_\pi)^2(u^3-2u)\Bigg\}\,, 
\\
{\rm Im}V_C &\!\!\!\!=\!\!\!\!& {g_A^2m_\pi^5\over (4f_\pi)^6\pi^2}(u^2-2)\bigg({1\over
u^2} -2\bigg)\Bigg\{2\sqrt{u^2-4}\Big[24c_1+c_2(u^2-4)
+ 6c_3(u^2-2)\Big]B(u) \nonumber \\ &&  +u\bigg[c_2 \bigg(8-{5u^2\over
3}\bigg)+
6c_3(2-u^2)-24c_1\bigg] \Bigg\} + {3 g_A^2m_\pi^5\over (2f_\pi )^4 u}(2-u^2)^3 \,\bar e_{14}\,,  
\\
{\rm Im}W_C &\!\!\!\!=\!\!\!\!&  - {c_1 m_\pi^5
\over(2f_\pi)^6
\pi^2}\bigg\{{3g_A^2+1 \over 8} \sqrt{u^2-4}\,(2-u^2)+\bigg({3g_A^2+1
\over u}
-2g_A^2\, u\bigg) B(u)\bigg\}
\nonumber \\
 && -  {c_2 m_\pi^5\over (2f_\pi)^6 \pi^2} \bigg\{{1\over 96}
\sqrt{u^2-4}\,\Big[7u^2-6-u^4+g_A^2(5u^2-6-2u^4)\Big]
+{1\over 4 u}(g_A^2u^2-1-g_A^2) B(u)\bigg\}
\nonumber \\
 && - {c_3 m_\pi^5\over (4f_\pi)^6\pi^2}
\Bigg\{{2\over 9} \sqrt{u^2-4}\bigg[3(7u^2-6-u^4)+4g_A^2\bigg({32 \over
u}-12
-20u+7u^2-u^4\bigg) 
\nonumber \\
 && +g_A^4\bigg(114-{512\over u}+368u-169u^2
+7u^4+{192\over u+2}\bigg)\bigg] 
\nonumber \\
 && +{16\over
3u}\Big[g_A^4(6u^4
-30u^2+35)+g_A^2(6u^2-8)-3\Big] B(u)\Bigg\}
\nonumber \\
&& - {c_4 g_A^2m_\pi^5\over (4f_\pi)^6 \pi^2}
\Bigg\{{2\over 9}
\sqrt{u^2-4}\,\bigg[30-{128\over u}+80u-13u^2-2u^4+g_A^2\bigg({512 \over u}
-114-368u 
\nonumber \\ 
&& +169u^2-7u^4-{192 \over u+2}\bigg)\bigg]
+{16\over 3u}\Big[5-3u^2+g_A^2(30u^2-35-6u^4)\Big] B(u)\Bigg\}\,.
\end{eqnarray}
Consistent with the calculation of the $\pi N$ amplitude in Ref.~\cite{KGE12}, we applied  
relations between LECs, such that only $\bar e_{14}$ and $\bar e_{17}$ remain in the final 
result.

\subsubsection{Relativistic corrections}
\label{sec_rel}

This group consists of diagrams with one vertex proportional to $c_i$
and one $1/M_N$ correction.
A few representative graphs are shown in Fig.~\ref{fig_dia4}(c).
Since in this investigation we count $Q/M_N \sim (Q/\Lambda_\chi)^2$, these relativistic 
corrections  are formally of order N$^4$LO. 
The result for this group of diagrams read in our sign-convention \cite{Kai01a}:
\begin{eqnarray} 
V_C & = & {g_A^2\, L(\tilde{\Lambda};q) \over 32 \pi^2 M_N f_\pi^4 } \left[ 
(6c_3-c_2) q^4 +4(3c_3-c_2-6c_1)q^2 m_\pi^2+6(2c_3-c_2)m_\pi^4- 24(2c_1+c_3)m_\pi^6 w^{-2} 
\right] \,,
\nonumber \\
\label{eq_4cMC}
\\
W_C &=& -{c_4 \over 192 \pi^2 M_N f_\pi^4 } 
\left[ g_A^2 (8m_\pi^2+5q^2) + w^2 \right] q^2 \,  L(\tilde{\Lambda};q)
\,, \\
W_T  &=&  -{1\over q^2} W_S = {c_4 \over 192 \pi^2 M_N f_\pi^4 } 
\left[ w^2-g_A^2 (16m_\pi^2+7q^2) \right]  L(\tilde{\Lambda};q)
\label{eq_4cMS}
\,,  \\
V_{LS}& = & {c_2 \, g_A^2 \over 8 \pi^2 M_N f_\pi^4 } 
\, w^2 L(\tilde{\Lambda};q) 
\,, \\
W_{LS}  &=& 
-{c_4  \over 48 \pi^2 M_N f_\pi^4 } 
\left[ g_A^2 (8m_\pi^2+5q^2) + w^2 \right]  L(\tilde{\Lambda};q)
\,,
\label{eq_4cMLS}
\end{eqnarray}
where the (regularized) logarithmic loop function is given by:
\begin{equation} 
L(\tilde{\Lambda};q) = {w\over 2q} 
\ln {\frac{\tilde{\Lambda}^2(2m_\pi^2+q^2)-2m_\pi^2 q^2+\tilde{\Lambda}\sqrt{
\tilde{\Lambda}^2-4m_\pi^2}\, q\,w}{2m_\pi^2(\tilde{\Lambda}^2+q^2)}}
\label{eq_L}
\end{equation}
with $ w = \sqrt{4m_\pi^2+q^2}$. Note that
\begin{equation}
\lim_{\tilde{\Lambda} \rightarrow \infty} L(\tilde{\Lambda};q) =  {w\over q} 
\ln {\frac{w+q}{2m_\pi}} \,,
\end {equation}
is the logarithmic loop function of dimensional regularization.

\subsection{Three-pion exchange contributions at N$^4$LO}
\label{sec_3pi}

\begin{figure}
\scalebox{0.9}{\includegraphics{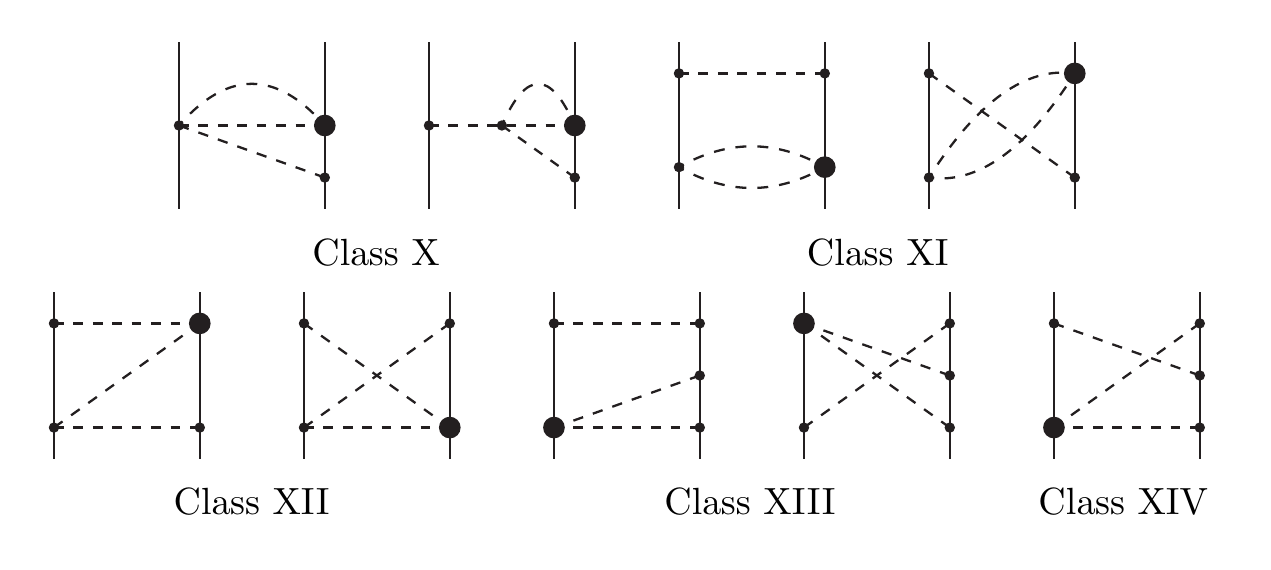}}
\caption{Three-pion exchange contributions at N$^4$LO.
The classification scheme of Ref.~\cite{Kai01} is used.
Notation as in Fig.~\ref{fig_dia4}.}
\label{fig_dia5}
\end{figure}

The $3\pi$-exchange of order N$^4$LO is shown in Fig.~\ref{fig_dia5}.
The spectral functions for these diagrams have been calculated in Ref.~\cite{Kai01}.
We use here the classification scheme introduced in that reference and note that class XI
vanishes. Moreover, we find that the class X and part of class XIV make only 
negligible contributions. Thus, we include in our calculations only class XII and XIII, and 
the $V_S$ contribution of class XIV.
In Ref.~\cite{Kai01} the spectral functions were presented in terms of an 
integral over the invariant mass of a pion pair. We have solved these integrals 
analytically and obtain the following spectral functions for the non-negligible cases:
\begin{eqnarray}
\operatorname{Im}V_S^{(\rm XII)} &=& -\frac{g_A^2 c_4m_\pi^5}{(4f_\pi)^6\pi^2u^3} \bigg[ 
\frac{y}{12}(u-1)(100u^3-27-50u-151u^2+185 u^4-14u^5-7u^6)
\nn && +4\,D(u)\,(2+10u^2-9u^4) \bigg] \,,
\\
\operatorname{Im}V_T^{(\rm XII)} &=& \frac{1}{\mu^2}\operatorname{Im}V_S^{(\rm XII)} - 
\frac{g_A^2c_4m_\pi^3}{(4f_\pi)^6\pi^2u^5} \bigg[ \frac{y}{6} (u-1) (u^6+2u^5-39u^4
-12u^3+65u^2-50u-27)
\nn &&+8 \,D(u)\,(3u^4-10u^2+2) \bigg] \,,
\\
\operatorname{Im}W_S^{(\rm XII)} &=& -\frac{g_A^2 m_\pi^5}{(4f_\pi)^6\pi^2u^3} \bigg\{ \,y\,
(u-1)\bigg[ \frac{4c_1u}{3} \bigg(u^3+2u^2-u+4 \bigg)
+\frac{c_2}{72} \bigg(u^6+2u^5-39u^4-12u^3+65u^2-50u-27 \bigg)
\nn &&
+\frac{c_3}{12} \bigg(u^6+2u^5-31u^4+4u^3+57u^2-18u-27 \bigg)
\nn &&
+\frac{c_4}{72} \bigg(7u^6+14u^5-185u^4-100u^3+151u^2+50u+27 \bigg) \bigg]
\nn &&
+\,D(u)\, \bigg[ 16c_1(4u^2-1-u^4)
+\frac{2c_2}{3} \bigg(2-10u^2+3u^4 \bigg)+4c_3 u^2(u^2-2)+\frac{2c_4}{3} \bigg( 9u^4-10u^2-2 
\bigg) \bigg] \bigg\} \,,
\end{eqnarray}
\begin{eqnarray}
\operatorname{Im}W_T^{(\rm XII)} &=& \frac{1}{\mu^2}\operatorname{Im}W_S^{(\rm XII)} - 
\frac{g_A^2 m_\pi^3}{(4f_\pi)^6\pi^2u^5}\bigg\{\,y\,(u-1) \bigg[ \frac{16c_1u}{3} \bigg( 
2+u-2u^2-u^3 \bigg)\nn &&
+\frac{c_2}{36} \bigg( 73u^4-6u^5-3u^6+44u^3-43u^2-50u-27 \bigg)
\nn &&
+\frac{c_3}{2} \bigg( 19u^4-2u^5-u^6+4u^3-9u^2-6u-9 \bigg)
\nn &&
+\frac{c_4}{36} \bigg( 39u^4-2u^5-u^6+12u^3-65u^2+50u+27 \bigg) \bigg]
\nn &&
+4\,D(u)\, \bigg[ 8c_1(u^4-1)+c_2 \bigg( \frac23-u^4 \bigg)
-2c_3 u^4+\frac{c_4}{3} \bigg(10u^2-2-3u^4 \bigg) \bigg] \bigg\} \,,
\end{eqnarray}
\begin{eqnarray}
\operatorname{Im}W_C^{(\rm XIII)} &=& -\frac{g_A^4c_4m_\pi^5}{(4f_\pi)^6\pi^2} \bigg[ 
\frac{8y}{3} (u-1)(u-4-2u^2-u^3) 
+32\,D(u)\, \bigg(u^3-4u+\frac1u \bigg) \bigg] \,, 
\\
\operatorname{Im}V_S^{(\rm XIII)} &=& -\frac{g_A^4c_4 m_\pi^5}{(4f_\pi)^6\pi^2u^3} \bigg[ 
\frac{y}{24} (u-1)(37u^6+74u^5-251u^4-268u^3+349u^2-58u-135)
\nn && +2\,D(u)\,(39u^4-2-52u^2-6u^6) \bigg] \,,
\\
\operatorname{Im}V_T^{(\rm XIII)} &=& \frac{1}{\mu^2}\operatorname{Im}V_S^{(\rm XIII)} - 
\frac{g_A^4c_4m_\pi^3}{(4f_\pi)^6\pi^2u^5} \bigg[ \frac{y}{12} (u-1)(5u^6+10u^5-3u^4
-252u^3-443u^2-58u-135)\nn && +4\,D(u)\,(3u^4+22u^2-2) \bigg] \,,
\\
\operatorname{Im}W_S^{(\rm XIII)} &=& -\frac{g_A^4m_\pi^5}{(4f_\pi)^6\pi^2u^3} \bigg\{ \,y\,
(u-1)\bigg[ 2c_1 u(5u^3+10u^2-5u-4)\nn &&
+\frac{c_2}{48} \bigg(135+58u-277u^2-36u^3+147u^4
-10u^5-5u^6 \bigg)
\nn &&
+\frac{c_3}{8} \bigg(7u^6+14u^5-145u^4-20u^3
+111u^2+18u+27 \bigg)
\nn &&
+\frac{c_4}{6} \bigg( 44u^3+37u^4-14u^5-7u^6-3u^2-18u-27 \bigg) \bigg]
\nn &&
+\,D(u)\, \bigg[ 24c_1(1+4u^2-3u^4)+c_2(2+2u^2-3u^4)
+6c_3 u^2(3u^2-2)+8c_4 u^2(u^4-5u^2+5) \bigg] \bigg\} \,,
\nn
\\
\operatorname{Im}W_T^{(\rm XIII)} &=& \frac{1}{\mu^2}\operatorname{Im}W_S^{(\rm XIII)} - 
\frac{g_A^4 m_\pi^3}{(4f_\pi)^6\pi^2u^5} \bigg\{ \,y\,(u-1) \bigg[ 4c_1u(5u^3+10u^2+7u-4)
\nn &&
+\frac{c_2}{24} \bigg(135+58u+227u^2+204u^3+27u^4-10u^5-5u^6 \bigg)
\nn &&
+\frac{c_3}{4}\bigg( 27+18u-9u^2-68u^3-121u^4
+14u^5+7u^6 \bigg)
\nn &&
+c_4(4u^3+19u^4-2u^5-u^6-9u^2-6u-9) \bigg]
\nn &&
+2 \,D(u)\, \bigg[ 24c_1(1-3u^4)
+c_2(2-10u^2-3u^4)+6c_3 u^2(3u^2+2)-8c_4 u^4 \bigg] \bigg\} \,,
\end{eqnarray}
\begin{eqnarray}
\operatorname{Im}V_S^{(\rm XIV)} &=& -\frac{g_A^4c_4 m_\pi^5}{(4f_\pi)^6\pi^2u^3} \bigg[ 
\frac{y}{24}(u-1)(637u^2-58u-135+116u^3-491u^4-22u^5-11u^6)
\nn &&
+2\,D(u)\,(6u^6-9u^4+8u^2-2) \bigg] \,,
\end{eqnarray}
where $y = \sqrt{(u-3)(u+1)}$ and $D(u) = \ln[(u-1+\,y)/2]$ with $u =\mu/m_\pi >3$.

\section{Perturbative $NN$ scattering in peripheral partial waves}
\label{sec_pertNN}

Nucleon-nucleon scattering in peripheral partial waves
is of special interest---for several reasons.
First, these partial waves probe the long- and 
intermediate-range of the nuclear force. Due to the centrifugal
barrier, there is only small sensitivity to short-range
contributions and, in fact, the contact terms up to and including order N$^3$LO  
make no contributions for orbital angular momenta $L\geq 3$.
Thus, for $F$ and higher waves and energies below the pion-production
threshold, we have a window in which the $NN$ interaction
is governed by chiral symmetry alone (chiral one- and multi-pion exchanges), and we can 
conduct a relatively clean test of how well the theory works.
Using values for the LECs from $\pi N$ analysis,
the $NN$ predictions are even parameter free.
Moreover, the smallness of the phase shifts in peripheral
partial waves suggests that the calculation can
be done perturbatively. This avoids the complications
and possible model-dependence (e.g., cutoff dependence)
that the non-perturbative treatment of the
Lippmann-Schwinger equation, necessary for low partial
waves, is beset with.
A thorough investigation of this kind at N$^3$LO was conducted in Ref.~\cite{EM02}.
Here, we will work at N$^4$LO.

The perturbative $K$-matrix for $np$ scattering
is calculated as follows:
\begin{eqnarray}
K({\vec p}~',\vec p) &=& 
 V_{1\pi}^{(np)} ({\vec p}~', \vec p) +
 V_{2\pi, \rm it}^{(np)}
({\vec p}~',{\vec p}) +
 V({\vec p}~',{\vec p}) 
\label{eq_kmat}
\end{eqnarray}
with $V_{1\pi}^{(np)} ({\vec p}~', \vec p)$ as in Eq.~(\ref{eq_1penp}), and
$V_{2\pi, \rm it}^{(np)} ({\vec p}~',{\vec p})$ representing the once iterated
one-pion exchange (1PE) given by
\begin{equation}
V_{2\pi, \rm it}^{(np)} ({\vec p}~',{\vec p})  = {\cal P}\!\!\int d^3p'' \:
\frac{M_N^2}{E_{p''}} \:
\frac{V_{1\pi}^{(np)}({\vec p}~',{\vec p}~'')\,
V_{1\pi}^{(np)}({\vec p}~'',{\vec p})} 
{{ p}^{2}-{p''}^{2}}
\,,
\label{eq_2piit}
\end{equation}
where ${\cal P}$ denotes the principal value integral and $E_{p''}=\sqrt{M_N^2+{p''}^2}$.
A calculation at LO includes only the first term on the right hand side of 
Eq.~(\ref{eq_kmat}), $ V_{1\pi}^{(np)} ({\vec p}~', \vec p) $, while calculations at NLO 
or higher order also include the second term on the right hand side,
$V_{2\pi, \rm it}^{(np)} ({\vec p}~',{\vec p})$.
At N$^3$LO and beyond, the twice iterated 1PE should be included, too.
However, we found that the difference between the once iterated 1PE and the 
infinitely iterated 1PE is so small that it could not be identified on the scale
of our phase shift figures. For that reason, we omit iterations of 1PE beyond what is
contained in $V_{2\pi, \rm it}^{(np)} ({\vec p}~',{\vec p})$.

Finally, the third term on the r.h.s. of Eq.~(\ref{eq_kmat}), $ V({\vec p}~',{\vec p}) $,
stands for the irreducible multi-pion exchange contributions that occur at the order at
which the calculation is conducted.
In multi-pion exchanges,
we use the average pion mass $m_\pi = 138.039$ MeV and, thus,
neglect the charge-dependence due to pion-mass splitting
in irreducible multi-pion diagrams.
The charge-dependence that emerges from irreducible $2\pi$ 
exchange was investigated in Ref.~\cite{LM98b} and found to be
negligible for partial waves with $L\geq 3$.

Throughout this paper, we use
\beq
M_N  =  \frac{2M_pM_n}{M_p+M_n} = 938.9182 \mbox{ MeV.}
\eeq
 Based upon relativistic kinematics,
the CMS on-shell momentum $p$ is related to
the kinetic energy of the incident neutron 
in the laboratory system (``Lab.\ Energy''), $T_{\rm lab}$, by
\beq
p^2  =  \frac{M_p^2 T_{\rm lab} (T_{\rm lab} + 2M_n)}
               {(M_p + M_n)^2 + 2T_{\rm lab} M_p}  
\,,
\eeq
with $M_p=938.2720$ MeV and $M_n=939.5653$ MeV
 the proton and neutron masses, respectively.

The $K$-matrix, Eq.~(\ref{eq_kmat}), is decomposed into partial waves following 
Ref.~\cite{EAH71}
and phase shifts are then calculated via
\begin{equation}
\tan \delta_L (T_{\rm lab}) = -\frac{M_N^2p }{16\pi^2E_p} \, p \, K_L(p,p)
\,.
\end{equation}
For more details concerning the evaluation of phase shifts, including the case of coupled 
partial waves, see Ref.~\cite{Mac93} or the appendix of \cite{Mac01}. All phase shifts
shown in this paper are in terms of Stapp conventions~\cite{SYM57}.

\begin{figure*}
\vspace*{-1cm}
\hspace*{-0.8cm}
\scalebox{0.75}{\includegraphics{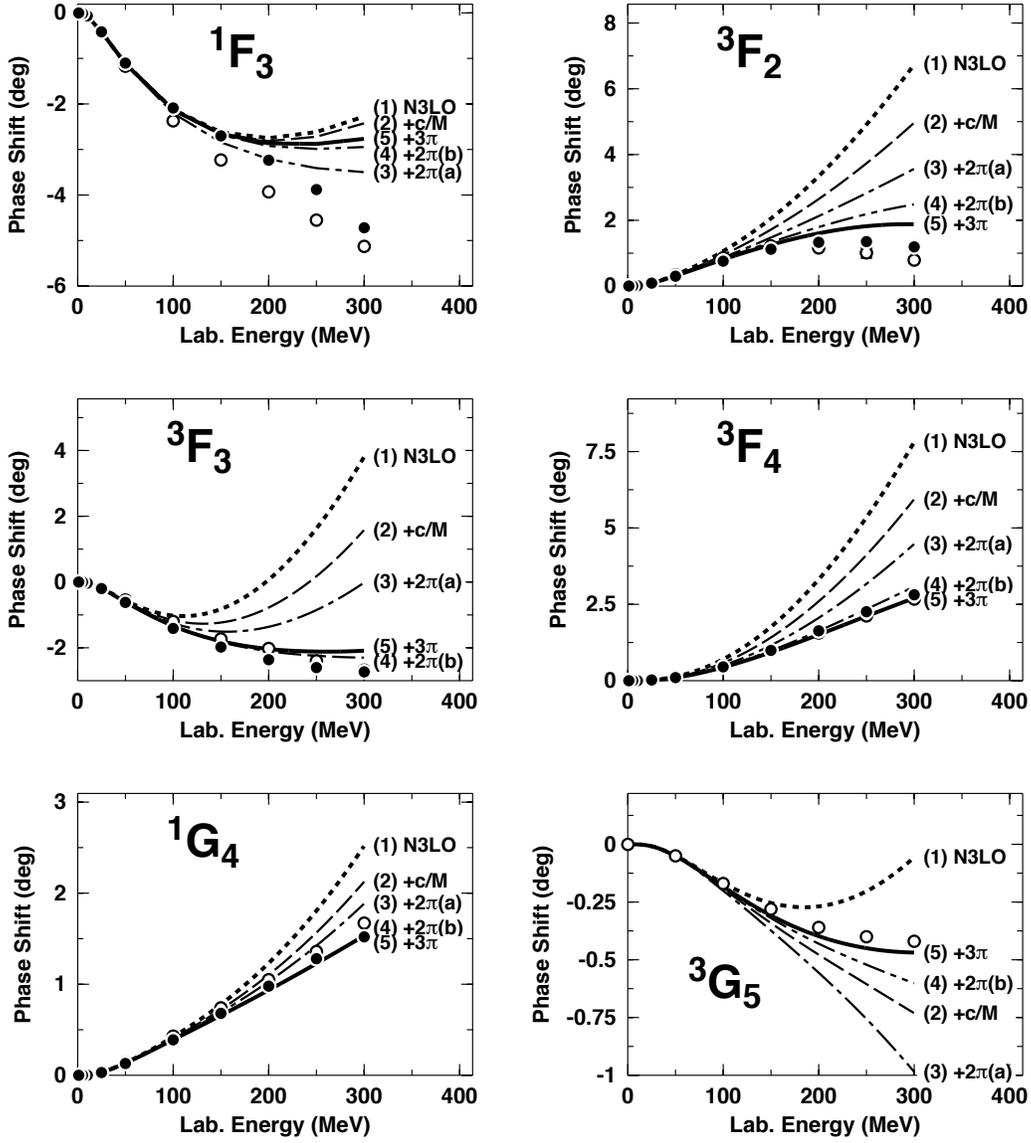}}
\vspace*{-2.5cm}
\caption{Effect of individual fifth-order contributions on the neutron-proton phase shifts 
of some selected peripheral partial waves. 
The individual contributions are added up successively in the order given in parenthesis
next to each curve. Curve (1) is N3LO and curve (5) is the complete N4LO.
The filled and open circles represent the results from the Nijmegan multi-energy $np$ phase-shift analysis~\cite{Sto93} and the VPI/GWU single-energy $np$ analysis SM99~\cite{SM99}, respectively.
\label{fig_ph1}}
\end{figure*}

We calculate phase shifts for
partial waves with $L\geq 3$ and 
$T_{lab}\leq 300$ MeV.
To establish a link between $\pi N$ and $NN$
and to check the consistency of the $\pi N$ and $NN$ systems, 
we use the $\pi N$ LECs determined in Ref.~\cite{KGE12}
in a calculation of
 $\pi N$ scattering at fourth order applying the same power counting scheme
as in the present work. To be specific, we use the set of LECs denoted by `KH' in Ref.~\cite{KGE12}. The values are:
\\ \\
$c_1=-0.75$ GeV$^{-1}$,  \hspace*{.3cm}
$c_2=3.49$ GeV$^{-1}$,  \hspace*{.3cm}
$c_3=-4.77$ GeV$^{-1}$,  \hspace*{.3cm}
$c_4=3.34$ GeV$^{-1}$;
\\ \\
$\bar{d}_1+\bar{d}_2=6.21$ GeV$^{-2}$,  \hspace*{.3cm}
$\bar{d}_3=-6.83$ GeV$^{-2}$,  \hspace*{.3cm}
$\bar{d}_5=0.78$ GeV$^{-2}$,  \hspace*{.3cm}
$\bar{d}_{14}-\bar{d}_{15}=-12.02$ GeV$^{-2}$;
\\ \\
$\bar{e}_{14}=1.52$ GeV$^{-3}$,  \hspace*{.3cm}
$\bar{e}_{17}=-0.37$ GeV$^{-3}$.
\\ \\
Moreover, we absorb the Goldberger-Treiman discrepancy
into an effective value for $g_A$, namely, $g_A=1.29$. Finally, the physical value of 
the pion-decay constant is $f_\pi=92.4$ MeV.

As shown in Figs.~\ref{fig_dia4} and \ref{fig_dia5} and derived in Sec.~\ref{sec_pions}, the fifth order consists of several contributions.
We will now demonstrate how the individual fifth-order contributions
impact $NN$ phase shifts in peripheral waves.
For this purpose, we display in 
Fig.~\ref{fig_ph1} 
phase shifts
for six important peripheral partial waves, namely,
$^1F_3$, $^3F_2$, $^3F_3$, $^3F_4$, $^1G_4$, and $^3G_5$.
In each frame, the following curves
are shown:
\begin{description}
\item[(1)]
N$^3$LO.
\item[(2)] The previous curve plus
the $c_i/M_N$ corrections (denoted by `c/M'),
Fig.~\ref{fig_dia4}(c) and Sec.~\ref{sec_rel}.
\item[(3)] The previous curve plus
the N$^4$LO $2\pi$-exchange (2PE) two-loop contributions of class (a),
Fig.~\ref{fig_dia4}(a) and Sec.~\ref{sec_2pia}.
\item[(4)] The previous curve plus
the N$^4$LO 2PE two-loop contributions of class (b),
Fig.~\ref{fig_dia4}(b) and Sec.~\ref{sec_2pib}.
\item[(5)] The previous curve plus
the N$^4$LO $3\pi$-exchange (3PE) contributions,
Fig.~\ref{fig_dia5} and Sec.~\ref{sec_3pi}.
\end{description}
In summary, the various curves add up
successively
the individual N$^4$LO contributions 
in the order indicated in the curve labels.
The last curve in this series, curve (5),
is the full N$^4$LO result.
In these calculations, a SFR cutoff $\tilde{\Lambda}=1.5$ GeV is applied [cf.\
Eq.~(\ref{eq_disp})].

From Fig.~\ref{fig_ph1}, we make the following observations. In triplet $F$-waves,
the $c_i/M_N$ corrections as well as the 2PE two-loops, class (a) and (b),
are all repulsive and of about the same strength. As a consequence, the problem of
the excessive attraction, that N$^3$LO is beset with, is overcome.
A similar trend is seen in $^1G_4$.
An exception is $^1F_3$, where the class (b) contribution is attractive leading to
phase shifts above the data for energies higher than 150 MeV. 

Now turning to the N$^4$LO 3PE contributions [curve (5) in Fig.~\ref{fig_ph1}]: 
they are substantially smaller than the 2PE two-loop ones, in all peripheral partial waves. 
This can be interpreted as an indication of convergence
with regard to the number of pions being exchanged between two nucleons---a trend
that is very welcome. Further, note that the total 3PE contribution is a very comprehensive one,
cf.\ Fig.~\ref{fig_dia5}. It is the sum of ten terms
(cf.\ Sec.~\ref{sec_3pi}) which, individually, can be fairly large.
However, destructive interference between them leads to the small net result.

For all $F$ and $G$ waves (except $^1F_3$), the final N$^4$LO result
is in excellent agreement with the empirical phase shifts. Notice that this includes
also $^3G_5$, which posed persistent problems at N$^3$LO~\cite{EM02}.

\begin{figure*}
\hspace*{-0.8cm}
\scalebox{0.75}{\includegraphics{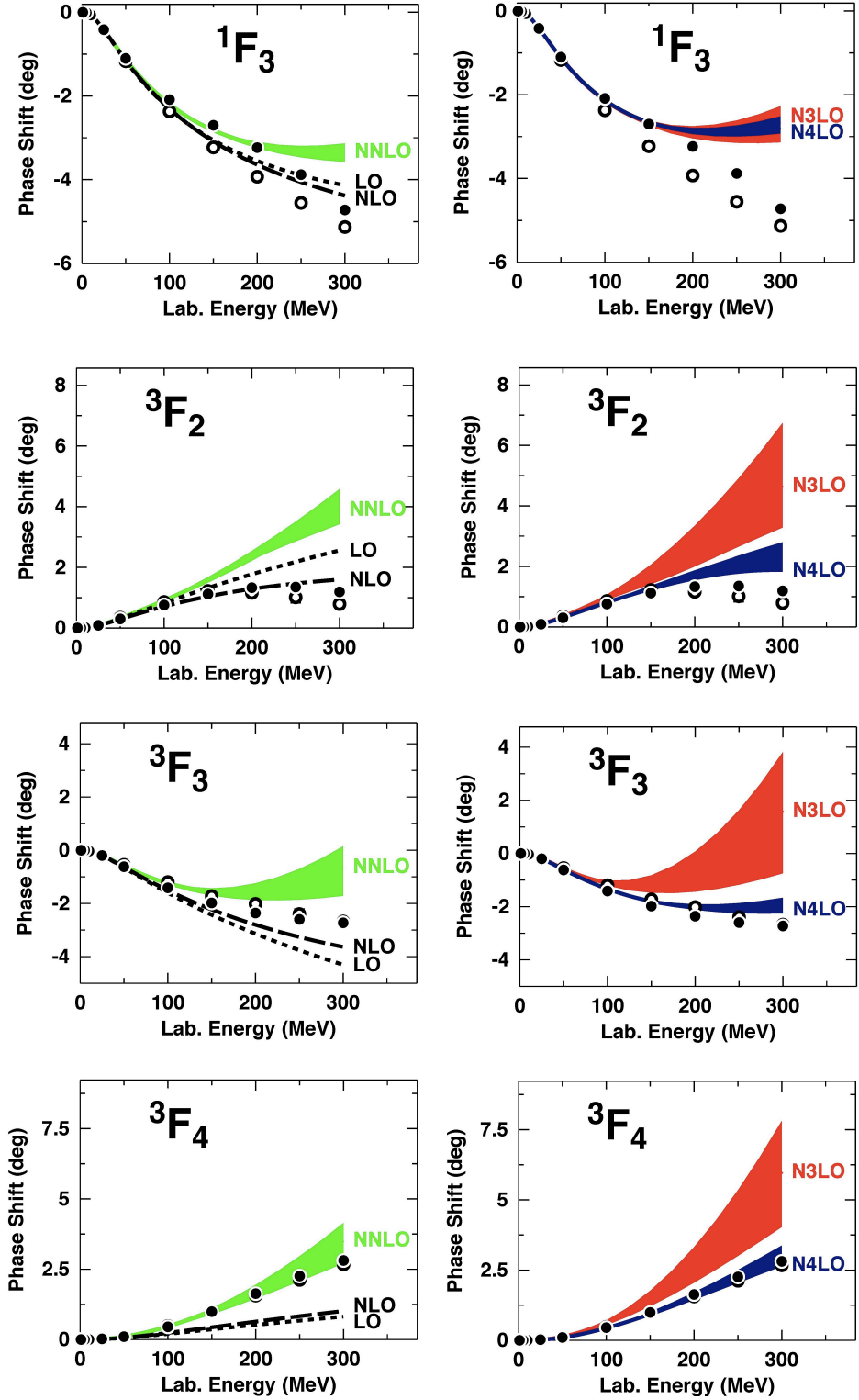}}
\caption{(Color online)
Phase-shifts of neutron-proton scattering at various orders as denoted. 
The shaded (colored) bands show the variation of the predictions when the
SFR cutoff $\tilde{\Lambda}$ is changed over the range 0.7 to 1.5 GeV.
The filled and open circles represent the results from the Nijmegan multi-energy $np$ phase-shift analysis~\cite{Sto93} and the VPI/GWU single-energy $np$ analysis SM99~\cite{SM99}, respectively.
\label{fig_ph5}}
\end{figure*}

\begin{figure*}
\hspace*{-0.8cm}
\scalebox{0.75}{\includegraphics{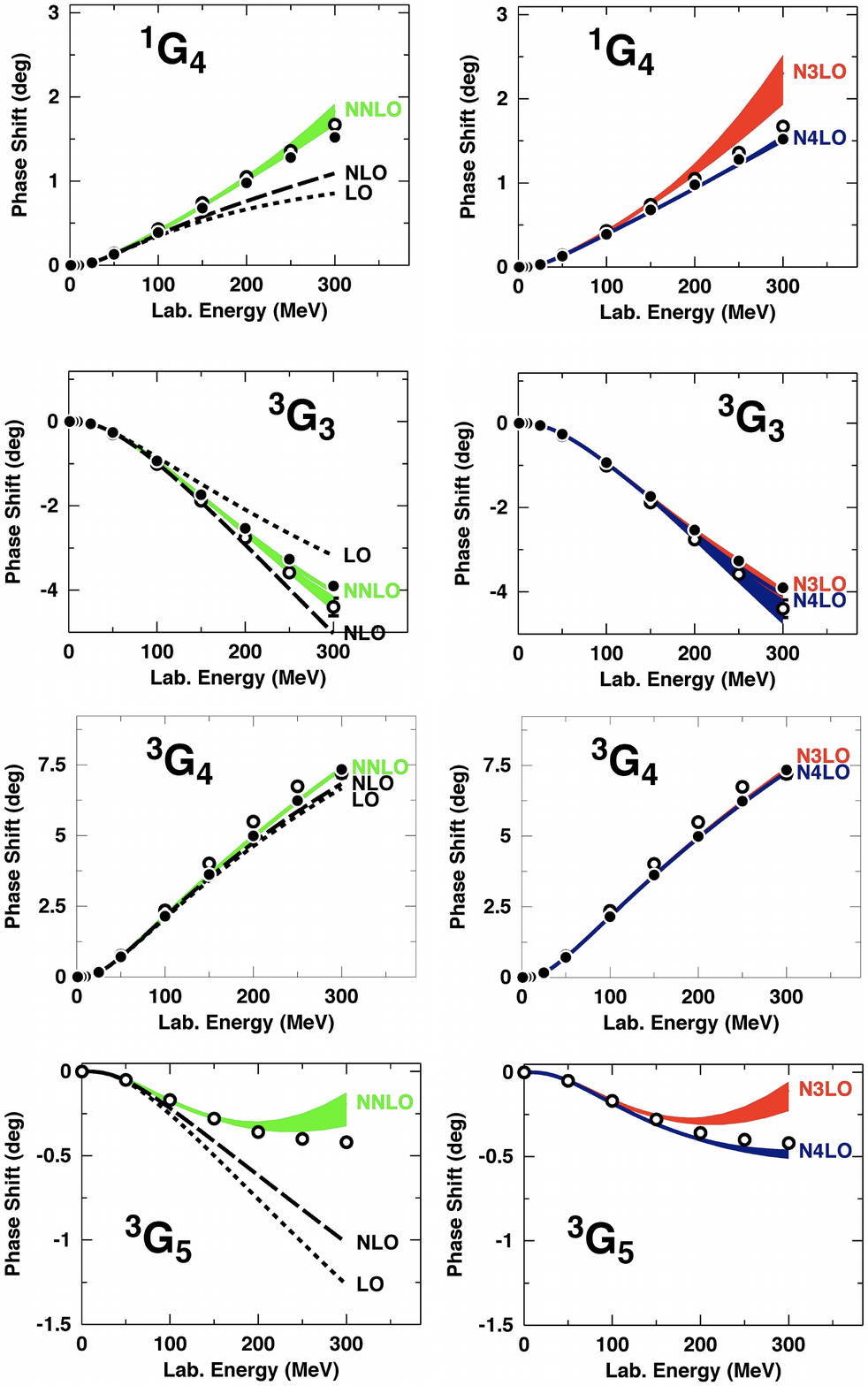}}
\caption{(Color online)
Same as Fig.~\ref{fig_ph5}, but for $G$-waves.
\label{fig_ph6}}
\end{figure*}

On a historical note, we mention that in the construction of the 
Stony Brook~\cite{JRV75,BJ76} and Paris~\cite{Vin79,Lac80} 
$NN$ potentials, which both include a 2PE contribution based upon dispersion theory,
the dispersion integral, Eq.~(\ref{eq_disp}), is cutoff at $\mu^2=50 \, m_\pi^2$, 
which is equivalent to a SFR cutoff $\tilde{\Lambda}= \sqrt{50} \, m_\pi \sim 1$ GeV.
Not accidentally, this agrees well with the common assumption of $\Lambda_\chi \sim 1$ GeV
and, thus,
sets the scale for an appropriate choice of  $\tilde{\Lambda}$. Consistent with this,
$\tilde{\Lambda}=1.5$ GeV was used for the results presented in Fig.~\ref{fig_ph1}.
It is, however, also of interest to know how predictions change with variations
of $\tilde{\Lambda}$ within a reasonable range. We have, therefore, varied $\tilde{\Lambda}$ between 0.7 and 1.5 GeV
and show the predictions for all $F$ and $G$ waves in Figs.~\ref{fig_ph5} and \ref{fig_ph6},
respectively, in terms of shaded (colored) bands. It is seen that, at N$^3$LO, the variations of the predictions
are very large and always too attractive while, at N$^4$LO, the variations are small
and the predictions are close to the data or right on the data.
 Figs.~\ref{fig_ph5} and \ref{fig_ph6} also include the lower orders 
 (as defined in the Appendices)
such that a comparison of the relative size of the order-by-order  contributions
is possible. We observe that there is not much of a convergence, since obviously the magnitudes
of the NNLO, N$^3$LO, and N$^4$LO contributions are about the same.
Potentially, this is characteristic for just these three orders and changes beyond N$^4$LO. 
But only an explicit calculation at N$^5$LO can settle this issue.

\section{Conclusions}
In this paper, we have calculated the one- and two-loop $2\pi$-exchange (2PE) and
two-loop $3\pi$-exchange (3PE) contributions to the $NN$ interaction which occur 
at N$^4$LO (fifth order) of the chiral low-momentum expansion.
The calculations are based upon heavy-baryon chiral perturbation theory
using the most general fourth order Lagrangian for pions and nucleons.
We apply $\pi N$ LECs, which were determined  in an analysis of elastic pion-nucleon
scattering to fourth order using the same power counting scheme as in the present work.
The spectral functions, which determine the $NN$ amplitudes via dispersion integrals,
are regularized by a cutoff $\tilde{\Lambda}$ in the range 0.7 to 1.5 GeV 
(also known as spectral-function regularization). Besides the cutoff $\tilde{\Lambda}$, 
our calculations do not involve any adjustable parameters.

From past work on $NN$ scattering in chiral perturbation theory (see, e.g., Ref.~\cite{EM02}), it is wellknown that, at NNLO and N$^3$LO, chiral 2PE produces far too much attraction.
The most important result of the present study is that
the N$^4$LO 2PE contributions are prevailingly repulsive and, thus,
compensate the excessive attraction of the lower orders.
As a consequence, the phase-shift predictions in $F$ and $G$ waves are in very good
agreement with the data, with the only exception of the $^1F_3$ wave.
The net 3PE contribution turns out to be moderate pointing towards convergence in terms of
the number of pions exchanged between two nucleons.
On the other hand, the NNLO, N$^3$LO, and N$^4$LO contributions are all about of the same
magnitude raising some concern about the convergence of the chiral
expansion of the $NN$ amplitude. To obtain more insight into this issue, future
investigations at N$^5$LO may be necessary.

\section*{Acknowledgements}
This work was supported in part by the U.S. Department of Energy
under Grant No.~DE-FG02-03ER41270 (R.M. and Y.N.),
the Ministerio de Ciencia y
Tecnolog\'\i a under Contract No.~FPA2010-21750-C02-02 and
the European Community-Research Infrastructure Integrating
Activity ``Study of Strongly Interacting Matter'' (HadronPhysics3
Grant No.~283286) (D.R.E.), and by DFG and NSFC (CRC110) (N.K.).

\appendix

\section{Leading order}
\label{sec_lo}
\begin{figure}
\scalebox{0.8}{\includegraphics{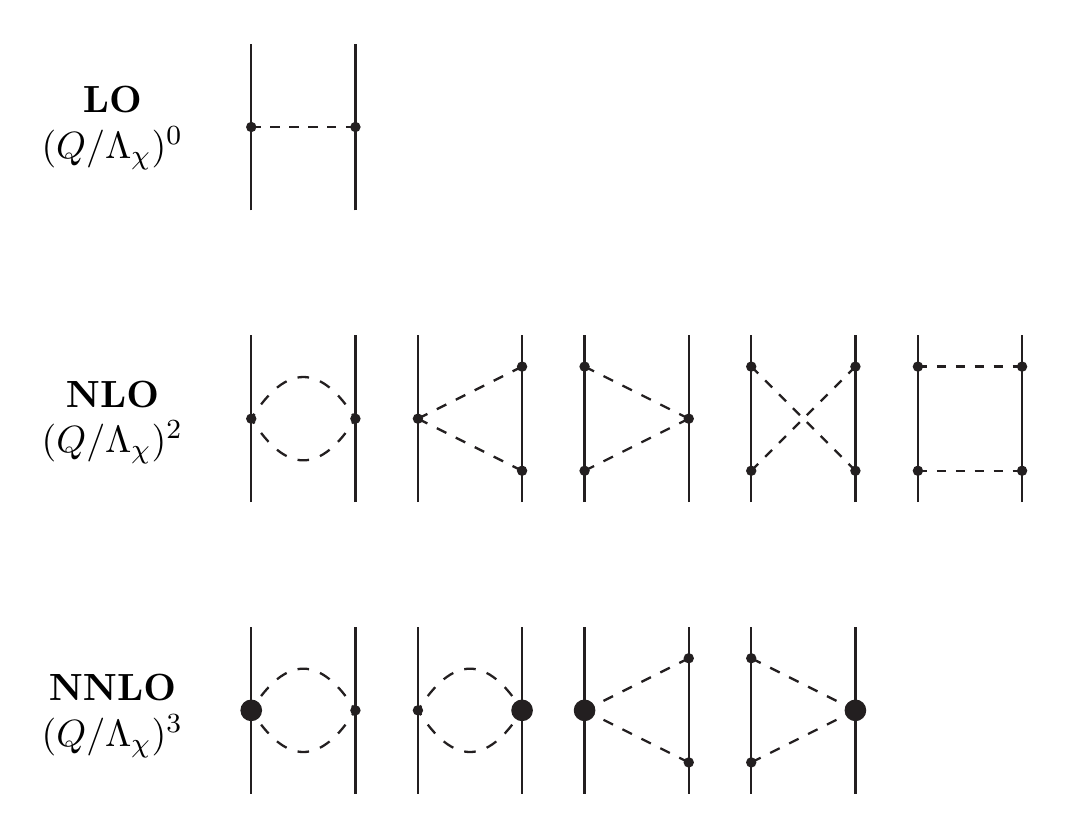}}
\caption{
LO, NLO, and NNLO contributions to the $NN$ interaction.
Notation as in Fig.~\ref{fig_dia4}.
\label{fig_dia1}}
\end{figure}

At leading order, there is only the $1\pi$-exchange contribution, cf.\ Fig.~\ref{fig_dia1}.
The charge-independent $1\pi$-exchange is given by
\begin{equation}
V_{1\pi}^{\rm(CI)} ({\vec p}~', \vec p) = - 
\frac{g_A^2}{4f_\pi^2}
\: 
\bm{\tau}_1 \cdot \bm{\tau}_2 
\:
\frac{
\vec \sigma_1 \cdot \vec q \,\, \vec \sigma_2 \cdot \vec q}
{q^2 + m_\pi^2} 
\,.
\label{eq_1PEci}
\end{equation}
Higher order corrections to the $1\pi$-exchange  are taken care of by  mass
and coupling constant renormalizations $g_A/f_\pi \to g_{\pi N}/M_N$. Note also that, on 
shell, there are no relativistic corrections. Thus, we apply  $1\pi$-exchange in the form
\eq{eq_1PEci} through all orders.

In this paper, we are specifically calculating neutron-proton ($np$) scattering 
and take the charge-dependence of the $1\pi$-exchange into account.
Thus, the $1\pi$-exchange  potential that we actually apply reads
\begin{equation}
V_{1\pi}^{(np)} ({\vec p}~', \vec p) 
= -V_{1\pi} (m_{\pi^0}) + (-1)^{I+1}\, 2\, V_{1\pi} (m_{\pi^\pm})
\,,
\label{eq_1penp}
\end{equation}
where $I=0,1$ denotes the total isospin of the two-nucleon system and
\begin{equation}
V_{1\pi} (m_\pi) \equiv - \,
\frac{g_A^2}{4f_\pi^2} \,
\frac{
\vec \sigma_1 \cdot \vec q \,\, \vec \sigma_2 \cdot \vec q}
{q^2 + m_\pi^2} 
\,.
\end{equation}
We use $m_{\pi^0}=134.9766$ MeV and
 $m_{\pi^\pm}=139.5702$ MeV.
Formally speaking, the charge-dependence of the 1PE exchange is of order 
NLO~\cite{ME11}, but we include it already at leading order to make the comparison with
the $np$ phase shifts more meaningful.

\section{Next-to-leading order}
\label{sec_nlo}

The $NN$ diagrams that occur at NLO (cf.\ Fig.~\ref{fig_dia1})
contribute in the following way~\cite{KBW97}:
\begin{eqnarray} 
W_C &=&{L(\tilde{\Lambda};q)\over384\pi^2 f_\pi^4} \left[4m_\pi^2(1+4g_A^2-5g_A^4)
+q^2(1+10g_A^2-23g_A^4) - {48g_A^4 m_\pi^4 \over w^2} \right] \,,  
\label{eq_2C}
\\   
V_T &=& -{1\over q^2} V_{S} \; = \; -{3g_A^4 \over 64\pi^2 f_\pi^4} L(\tilde{\Lambda};q)\,.
\label{eq_2T}
\end{eqnarray}

\section{Next-to-next-to-leading order}
\label{sec_nnlo}

The NNLO contribution (lower row of Fig.~\ref{fig_dia1}) is given by~\cite{KBW97}:
\begin{eqnarray} 
V_C &=&  {3g_A^2 \over 16\pi f_\pi^4} \left[2m_\pi^2(c_3- 2c_1)+c_3 q^2 \right](2m_\pi^2+q^2) 
A(\tilde{\Lambda};q) \,, \label{eq_3C}
\\
W_T &=&-{1\over q^2}W_{S} =-{g_A^2 \over 32\pi f_\pi^4} c_4 w^2  A(\tilde{\Lambda};q)\,.
\label{eq_3T}
\end{eqnarray}   
The loop function that appears in the above expressions,
regularized by spectral-function cut-off $\tilde{\Lambda}$, is
\begin{equation} 
A(\tilde{\Lambda};q) = {1\over 2q} \arctan{q (\tilde{\Lambda}-2m_\pi) \over q^2
+2\tilde{\Lambda} m_\pi} \,.\label{eq_A}
\end{equation}
Note that \begin{equation}
\lim_{\tilde{\Lambda} \rightarrow \infty} A(\tilde{\Lambda};q) =  
{1\over 2q} \arctan{q \over 2m_\pi} 
\end {equation}
yields the loop function used in dimensional regularization.

\section{Next-to-next-to-next-to-leading order}
\label{sec_n3lo}

\begin{figure}
\scalebox{0.80}{\includegraphics{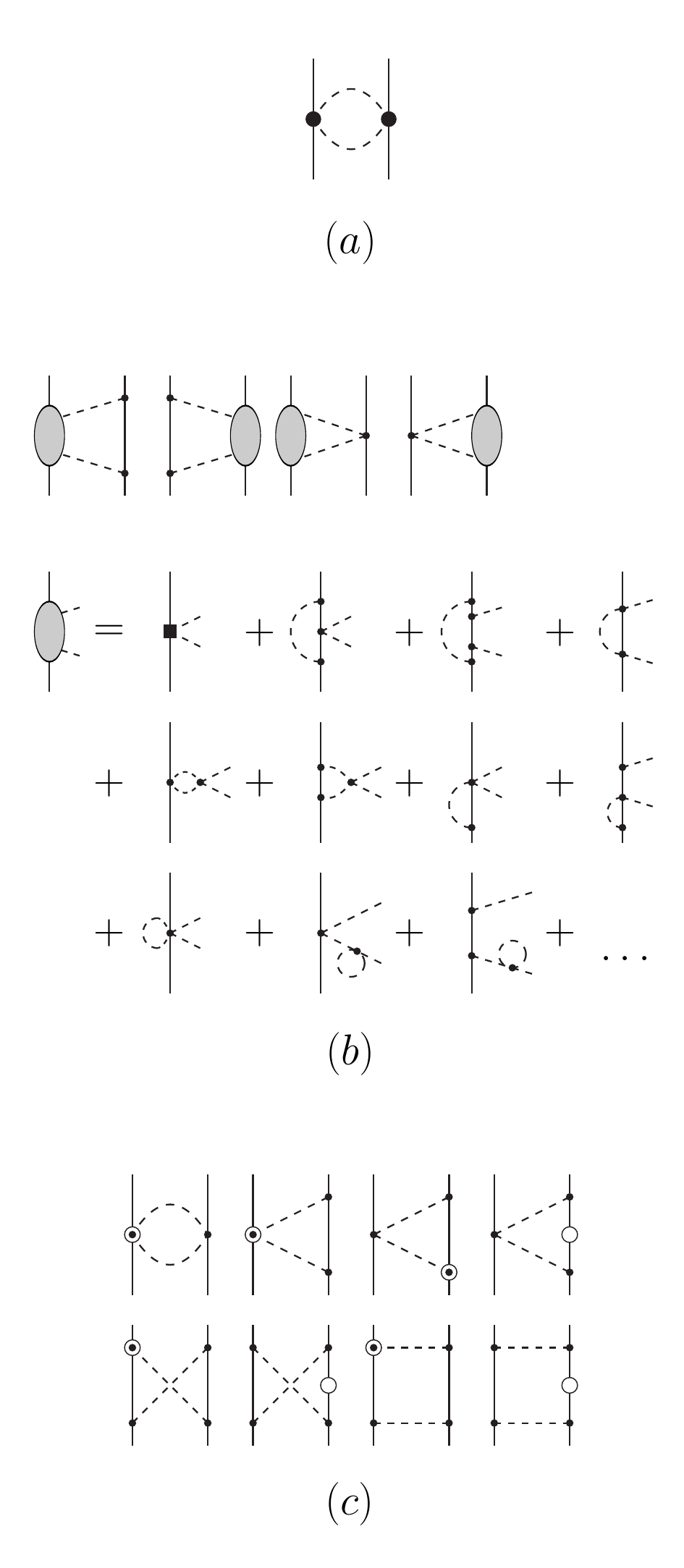}}
\caption{Two-pion exchange contributions at N$^3$LO
with (a) the N$^3$LO football diagram, (b) the leading 2PE two-loop contributions,
and (c) the relativistic corrections of NLO diagrams.
Notation as in Fig.~\ref{fig_dia4}.}
\label{fig_dia2}
\end{figure}

\subsection{Football diagram at N$^3$LO}

The football diagram at N$^3$LO, Fig.~\ref{fig_dia2}(a), generates~\cite{Kai01a}:
\begin{eqnarray}  
V_C & = & {3\over 16 \pi^2 f_\pi^4 } 
\left[\left( {c_2 \over 6} w^2 +c_3(2m_\pi^2+q^2) -4c_1 m_\pi^2 \right)^2 
+{c_2^2 \over 45 } w^4 \right]  L(\tilde{\Lambda};q) \,, 
\label{eq_4c2C}
\\
W_T  &=&  -{1\over q^2} W_S 
     = {c_4^2 \over 96 \pi^2 f_\pi^4 }  w^2 L(\tilde{\Lambda};q)
\,.
\label{eq_4c2T}
\end{eqnarray}

\subsection{Leading two-loop contributions}

The leading order $2\pi$-exchange two-loop diagrams are shown in Fig.~\ref{fig_dia2}(b).
In terms of spectral functions, the results are~\cite{Kai01a}:
\begin{eqnarray} 
{\rm Im}\, V_C &=& {3g_A^4 (2m_\pi^2-\mu^2) \over \pi \mu (4f_\pi)^6} \left[ (m_\pi^2-2\mu^2) 
\left( 2m_\pi +{2m_\pi^2 -\mu^2 \over2\mu} \ln{\mu+2m_\pi \over \mu-2m_\pi} \right) 
+4g_A^2 m_\pi(2m_\pi^2-\mu^2) \right] \,,
\\
{\rm Im}\, W_C &=& {2\kappa \over 3\mu (8\pi f_\pi^2)^3} \int_0^1 dx\, 
\Big[ g_A^2(\mu^2-2m_\pi^2) +2(1-g_A^2)\kappa^2x^2 \Big]
\nonumber \\ && \times \Bigg\{ \,96 \pi^2 f_\pi^2 \left[ (2m_\pi^2-\mu^2)(\bar{d}_1 
+\bar{d}_2) -2\kappa^2x^2 \bar{d}_3+4m_\pi^2 \bar{d}_5 \right] 
\nonumber \\ && +\left[ 4m_\pi^2 (1+2g_A^2) -\mu^2(1+5g_A^2)\right] 
{\kappa\over \mu} \ln {\mu +2\kappa\over 2m_\pi} \,
+\,{\mu^2 \over 12} (5+13g_A^2) -2m_\pi^2 (1+2g_A^2) 
\nonumber \\ && 
-\,3\kappa^2x^2 +6 \kappa x \sqrt{m_\pi^2 +\kappa^2 x^2} \ln{ \kappa x +\sqrt{m_\pi^2 
+\kappa^2 x^2}\over  m_\pi}
\nonumber \\ && +g_A^4\left(\mu^2 -2\kappa^2 x^2 -2m_\pi^2\right) 
\left[ {5\over 6} +{m_\pi^2\over \kappa^2 x^2} 
-\left( 1 +{m_\pi^2\over \kappa^2 x^2} \right)^{3/2} 
\ln{ \kappa x +\sqrt{m_\pi^2 +\kappa^2 x^2}\over  m_\pi} \right] \Bigg\} \,,   
\\
{\rm Im}\, V_S &=&\mu^2\,{\rm Im}\, V_T = 
{g_A^2\mu \kappa^3 \over 8\pi f_\pi^4} \left(\bar{d}_{15}-\bar{d}_{14}\right) 
\nonumber \\ &&+{2g_A^6\mu \kappa^3 \over (8\pi f_\pi^2)^3} 
\int_0^1 dx(1-x^2)\left[ {1\over 6}-{m_\pi^2 \over \kappa^2x^2} 
+\left( 1+{m_\pi^2 \over \kappa^2x^2} \right)^{3/2} 
\ln{ \kappa x +\sqrt{m_\pi^2 +\kappa^2 x^2}\over  m_\pi}
\right] \,,
\\
{\rm Im}\, W_S &=& \mu^2 \,{\rm Im}\, W_T(i\mu) = 
{g_A^4(4m_\pi^2-\mu^2) \over \pi (4f_\pi)^6} \left[ \left( m_\pi^2 -{\mu^2 \over 4} \right)
\ln{\mu+2m_\pi \over \mu-2m_\pi} 
+(1+2g_A^2)\mu  m_\pi\right]\,, 
\end{eqnarray}
where $\kappa = \sqrt{\mu^2/4-m_\pi^2}$.

The momentum space amplitudes $V_\alpha(q)$ and $W_\alpha(q)$
are obtained from
the above expressions by means of the dispersion integrals shown in Eq.~(\ref{eq_disp}).

\subsection{Leading relativistic corrections}

The relativistic corrections of the NLO diagrams, which are shown in Fig.~\ref{fig_dia2}(c),
count as N$^3$LO and are given by~\cite{ME11}:
\begin{eqnarray}
V_C &=& \frac{3 g_A^4}{128 \pi f_\pi^4 M_N} 
\bigg[\frac{m_\pi^5}{2w^2}+(2m_\pi^2+q^2)(q^2-m_\pi^2) A(\tilde{\Lambda};q) \bigg]
\,,
\label{eq_3EM1}
\\
W_C &=& \frac{g_A^2}{64 \pi f_\pi^4 M_N} 
\Bigg\{ \frac{3g_A^2m_\pi^5}{2\omega^2} +\big[g_A^2 (3m_\pi^2+2q^2) - 2m_\pi^2-q^2\big] 
(2m_\pi^2+q^2) A(\tilde{\Lambda};q) \Bigg\}
\,,
\\
V_T &=& -\frac{1}{q^2} V_S = \frac{3 g_A^4}{256 \pi f_\pi^4 M_N} 
(5m_\pi^2+2q^2) A(\tilde{\Lambda};q)\,,
\\
W_T &=& -\frac{1}{q^2} W_S = \frac{g_A^2}{128 \pi f_\pi^4 M_N} 
\big[g_A^2 (3m_\pi^2+q^2)-w^2 \big] A(\tilde{\Lambda};q) \,,
\label{eq_3EM4}
\\
V_{LS} &=&  {3g_A^4  \over 32\pi f_\pi^4 M_N} \, (2m_\pi^2+q^2) A(\tilde{\Lambda};q)
 \,,\\  
W_{LS} &=& {g_A^2(1-g_A^2)\over 32\pi f_\pi^4 M_N} \, w^2 A(\tilde{\Lambda};q) \,.
\end{eqnarray}

\subsection{Leading three-pion exchange contributions}
The leading $3\pi$-exchange contributions that occur at N$^3$LO 
have been calculated in Refs.~\cite{Kai00a,Kai00b} and are found to be negligible. We, 
therefore, omit them.

\end{document}